\documentclass[aps,pra,twocolumn,preprint,10pt]{revtex4-1} 
\usepackage{amsmath,amssymb,graphicx}
\usepackage[colorlinks]{hyperref}
\hypersetup{
 colorlinks=true,
 citecolor=blue,
 linkcolor=black,
 urlcolor=black}
\usepackage{breakurl}
\newcommand{\ket}[1]{|{#1}\rangle}
\newcommand{\Hi}{{\mathcal{H}}}
\newcommand{\bra}[1]{\langle{#1}|}

\newcommand{\norm}[1]{\left|{#1}\right|}
\newcommand{\eq}[1]{(\ref{eq:#1})}
\renewcommand{\sec}[1]{\hyperref[sec:#1]{Section~\ref{sec:#1}}}
\newcommand{\app}[1]{\hyperref[sec:#1]{Appendix~\ref{sec:#1}}}
\newcommand{\fig}[1]{\hyperref[fig:#1]{FIG.~\ref{fig:#1}}}

\newcommand{\si}{({\bf I}) }
\newcommand{\sii}{({\bf II}) }
\begin{document}

\title{What is the best way to teleport a qudit?}

\author{Kevin Marshall}\email{Corresponding author: marshall@physics.utoronto.ca}
\affiliation{Department of Physics, University of Toronto, 60 St. George St.,
Toronto, Ontario, M5S 1A7, Canada}

\author{Daniel F. V. James}
\affiliation{Department of Physics, University of Toronto, 60 St. George St.,
Toronto, Ontario, M5S 1A7, Canada}

\begin{abstract}
The need for high fidelity quantum teleportation arises in a variety of quantum algorithms and protocols.  Unfortunately, conventional continuous variable teleportation schemes rely on EPR states that yield a fidelity that approaches unity only in the limit of an unphysical amount of squeezing.  A new method, which utilizes an ensemble of single photon entangled states to teleport continuous variable states with fidelity approaching unity with finite resources was recently proposed by Andersen and Ralph [Phys. Rev. Lett. 111, 050504 (2013)].  We extend these ideas to consider the general case of using maximally entangled states of an arbitrary dimension to teleport a continuous variable state and discuss how the corresponding results are affected.
\end{abstract}

\maketitle 
\section{Introduction}
Quantum teleportation \cite{tele1} allows one party to send an arbitrary quantum state to another using only a classical channel, local operations, and a preshared entanglement resource.  Such teleportation protocols can be classified as either discrete or continuous variable (CV) based on the nature of the state that is to be transferred.  Discrete teleportation deals with quantum states in a finite dimensional Hilbert space \cite{tele1,tele2}, such as the polarization of a photon \cite{expr1,expr2}, or the energy level of an atom \cite{blatt}.  Continuous variable teleportation deals with states in an infinite dimensional Hilbert space \cite{vaidman,cvtele}, such as the quadratures of a single mode of the EM field, or the position of an atom in a harmonic potential.  Quantum teleportation has proven to be a useful module in the field of quantum computation for both algorithm development as well as assisting in carrying out actual computations \cite{tele3,asymp,port,word}.  Continuous variable teleportation has the advantage that it can be carried out using relatively simple states, transformations and detectors \cite{ralph}.

The conventional method of CV teleportation relies on a two-mode squeezed vacuum state which serves as an entanglement resource between two parties.  Unfortunately, the teleportation fidelity is highly dependent on the amount of two-mode squeezing, and using state-of-the-art technology the highest fidelity measured to date is only $83\%$ \cite{cvlimit}.  Many important quantum protocols require about 10 dB of two-mode squeezing \cite{squeeze}, however achieving high fidelity is limited not only by technical details but it is not possible even in theory to attain 100\% fidelity as this would require infinite squeezing, which is unphysical.  In a recent paper Andersen and Ralph explore a new approach to the teleportation of continuous variable states \cite{ralph}.  In this protocol, one replaces the standard two-mode squeezed state resource with a set of maximally entangled states on two dimensional Hilbert spaces in order to attain teleportation fidelities close to 100\% with modest resources.  In this paper we explore an extension of these ideas to Hilbert spaces of arbitrary dimension, as well as discuss the possibility of using alternative intermediary teleportation protocols.

The advantage of this new scheme depends on the ability to teleport an optical state containing up to $d$ photons defining a qudit.  There are a number of different propositions for such generalized quantum scissor protocols, some of which will be discussed in later sections.  The advantage of moving to a higher dimension is that one will need to implement less teleportation modules, and if the modules are sufficiently simple this may result in a more robust protocol.

This paper is organized as follows, first we state the formalism used in the conventional teleportation of a state on a finite dimensional Hilbert space.  Then we consider a new scheme \sii where a CV state is teleported using a set of maximally entangled qudits, this is an extension of the scheme \si developed in Ref. \cite{ralph} where the qudits are taken to be of dimension two.  Next we demonstrate how this new protocol implements entanglement swapping, as well as considering sources of noise and loss.  Finally, we provide a discussion detailing how this new scheme might be implemented, how it compares to other protocols, and what advantages it might offer.
\section{Conventional Teleportation}
In the conventional CV teleportation protocol, there is a continuous degree of freedom in both the squeezing, and thus entanglement, of the shared resource as well as the nature of the local recovery operation that needs to be performed by the recipient.  It should not be surprising that by threading our state through a set of finite dimensional Hilbert spaces something must be lost.  Specifically, all of the higher dimensional terms will be ``cut-off'' when teleporting a state by using an entangled resource of a lower dimension \cite{scissors}.  So in order to teleport a CV state in this manner with high fidelity, we require extra information about the state to be teleported; we need to know the average photon number in order to ensure not too much information is lost upon truncation.

Let us review the traditional teleportation procedure on a finite dimensional Hilbert space \cite{tele2}.  This is of interest because it can be used as a building block in the process of teleporting a CV quantum state.  Suppose Alice wishes to transfer the quantum state of an arbitrary qudit $\ket{\phi}_1 \in \Hi$ with $dim(\Hi)=d$.  Alice and Bob share an entangled quantum state of the form
\begin{align}
\ket{\psi}_{23}&=\frac{1}{\sqrt d}\sum_{m=0}^{d-1}\ket{m}_2\ket{m}_3,
\end{align}
where Bob has system 3 while Alice has systems 1 and 2.  Define unitary operations consisting of
\begin{subequations}
\begin{align}
\label{eq:xor}
\widehat{XOR}_{21}\ket{j}_1\ket{i}_2&=\ket{i \ominus j}_1\ket{i}_2\\
\hat{Z}_3\ket{m}_1&=\omega^m\ket{m}_3\\
\hat{X}_3\ket{m}_1&=\ket{m\oplus 1}_3
\end{align}
\end{subequations}
where $\omega=\exp(2\pi i/d)$ and the symbols $\oplus,\ominus$ refer to addition and subtraction modulo $d$ respectively.  To teleport a state $\ket{\phi}_3$ Alice applies the XOR operator on her systems as follows:
\begin{align}
\widehat{XOR}_{21}\ket{\phi}_1\ket{\psi}_{23}&=\frac{1}{d}\sum_{\ell,k=0}^{d-1}\hat{Z}_3^{d-\ell}\hat{X}_3^k\ket{k}_1\ket{\nu_\ell}_2\ket{\phi}_3
\end{align}
where
\begin{align}
\ket{\nu_\ell}_2&=\frac{1}{\sqrt d}\hat{Z}_2^\ell\sum_{k=0}^{d-1}\ket{k}_2.
\end{align}
Notice that the $d^2$ operators of the form $\hat{Z}_3^{d-\ell}\hat{X}_3^k$ are in one to one correspondence with the joint states of the form $\ket{k}_1\ket{\nu_\ell}_2$, and thus a projection onto these orthonormal states with outcome corresponding to $\ket{k'}_1\ket{\nu_\ell'}_2$ would project system 3 onto the normalized state $\hat{Z}_3^{d-\ell'}\hat{X}_3^{k'}\ket{\phi}_3$.  Alice is then able to send Bob the classical information $(\ell',k')$, from which he can apply a correction operation corresponding to $(\hat{X}_3^{k'})^\dagger(\hat{Z}_3^{d-\ell'})^\dagger$.  If the quantum state is ideal and maximally entangled to begin with, the $\ket{\nu_\ell}$ states correspond to the quantum Fourier transformed basis states, a well known and studied unitary operation \cite{qft}.
\section{CV teleportation via qudits}
In order to use a set of maximally entangled states of the form
\begin{align}
\label{eq:psi}
\ket \psi &= \frac{1}{\sqrt d}\sum_{k=0}^{d-1}\ket k\ket k
\end{align}
to teleport a CV state, we must first divide the input state into a number of intermediate states; for the purposes of this paper we will consider the input state to be a mode of the EM field.  Perhaps the most intuitive approach is to evenly split the input mode into a set of $N$ modes by way of an array of $N-1$ asymmetric beamsplitters called an $N$-splitter \cite{nsplit}.  It will be convenient to first study how such a protocol will transform a coherent state, and then to use this information to deduce a generalization.  A coherent state can be expressed in terms of the Fock basis as

\begin{align}
\label{eq:coherent}
e^{-\norm \alpha^2/2}\sum_{n=0}^\infty \frac{\alpha^n}{n!}(\hat a^\dagger)^n\ket 0,
\end{align}
where $\ket 0$ denotes the vacuum state.  Coherent states are often thought of as semi-classical states, and it should not be surprising that a coherent state will transform through an $N$-splitter as $\ket \alpha \rightarrow \ket{\alpha/\sqrt N}^{\otimes N}$.  In other words, the state is evenly split amongst the $N$ intermediate modes.

\begin{figure}[htp]
\centering
\includegraphics{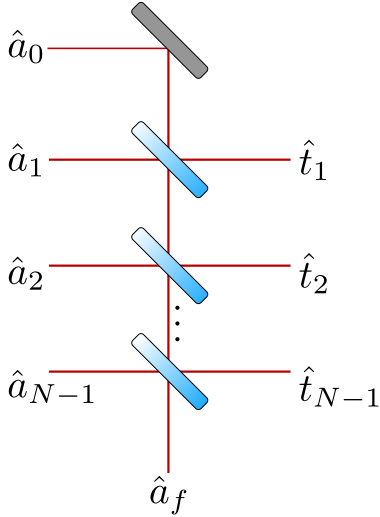}
\caption{Recombining $N$ modes on an $N$-splitter.}
\label{fig:ncombine}
\end{figure}

Although each of these modes $\ket{\alpha/\sqrt N}$ is still a coherent state in an infinite dimensional Hilbert space, the average number of photons $\bar n = \langle \hat a^\dagger\hat a\rangle$ is reduced by a factor $N$.  Equivalently, the coefficients associated with high photon numbers in \eq{coherent} have smaller amplitudes, and thus less information will be lost if we were to truncate the state at some suitable dimension.  Consider a truncation of the state after it is passed through the $N$-splitter given by
\begin{multline}
\label{eq:expansion}
\ket{\alpha/\sqrt N}^{\otimes N}\rightarrow\left[\exp\left(-\frac{\norm\alpha^2}{2N}\right)\left(\hat{\mathcal I} + \frac{\alpha}{\sqrt N}\hat a^\dagger +\right.\right. \\+ \left.\left.\frac{\alpha^2}{2N}(\hat a^\dagger)^2+\ldots + \frac{\alpha^d}{d!(\sqrt N)^d}(\hat a^\dagger)^d\right)\ket 0\right]^{\otimes N}
\end{multline}
where we have dropped all terms corresponding to photon number larger than $d$.  This result corresponds to what we would be left with if we were to implement some form of teleportation for each mode on a Hilbert space of dimension $d$+1.  The scheme developed by Andersen and Ralph \si ends this truncation at $d=1$, corresponding to keeping only the vacuum and one-photon state, the scheme we propose \sii is a generalization to arbitrary dimensions.  The final step of the teleportation protocol is to recombine the $N$ modes on an inverted $N$-splitter as shown in \fig{ncombine}.
If we condition on the outcome that all of the modes $t_1,t_2,\ldots,t_{N-1}$ contain no photons, then each of the annihilation operators in \eq{expansion} simply pick up another factor of $\frac{1}{\sqrt N}$ and become $\hat a_f$.  Thus we are left with the state
\begin{align}
\label{eq:phi}
\ket \phi &=\mathcal{N}\left(\hat{\mathcal I} + \frac{\alpha}{N}\hat a_f^\dagger +\ldots+ \frac{\alpha^d}{d!N^d}(\hat a_f^\dagger)^d\right)^{N}\ket{0},
\end{align}

where $\mathcal N$ is some normalization constant.

It will be useful to explicitly write $\ket \phi$ in the Fock basis as $\ket \phi=\sum_{k=0}^M c_k \ket k$ for some set of coefficients $c_k$.  The coefficient $c_k$ is given by:
\begin{align}
{N\brace k}_d~ \left(\frac{\alpha}{N}\right)^k \sqrt{k!}
\end{align}
where
\begin{align}
{N\brace k}_d &\equiv\sum_{\substack{\{r_1,r_2,\ldots,r_N\} \\ r_1+r_2+\ldots+r_n=k \\ r_i \leq d~ \forall i}} \prod_{r_j} \frac{1}{r_j!}
\end{align}
results from a counting argument; see \hyperref[sec:app]{Appendix}.  Using this we can write the final output as
\begin{align}
\ket \phi &= \frac{e^{-\norm\alpha^2/2}}{\sqrt{P_{suc}}}\sum_{k=0}^{dN} {N\brace k}_d \left(\frac{\alpha}{N}\right)^k \sqrt{k!} \ket k
\end{align}
where
\begin{align}
P_{suc}&=e^{-\norm\alpha^2} \sum_{k=0}^{dN}{N\brace k}_d^2 \left(\frac{\norm\alpha^2}{N^2}\right)^k k!
\end{align}
is the probability of detecting no photons in any of the $t$-modes.

Observe that an arbitrary Fock state is transformed through the protocol as
\begin{align}
\label{eq:trans}
\ket k &\rightarrow {N\brace k}_d \frac{k!}{N^k}\ket k,
\end{align}
so we have that an arbitrary state $\ket \Phi=\sum_{k=0}^\infty c_k \ket k$ transforms as
\begin{align}
\label{eq:tele}
\ket{\Phi_{tele}} &= \frac{1}{\sqrt{P_{suc}}}\sum_{k=0}^{dN} c_k {N\brace k}_d~ \frac{k!}{N^k}\ket k\\
\label{eq:psuc}
P_{suc}&=\sum_{k=0}^{dN} \norm{c_k}^2 {N\brace k}_d^2\frac{k!^2}{N^{2k}},
\end{align}
where $P_{suc}$ ensures normalization.  In the case that we limit ourselves to a two-dimensional Hilbert space, of only the vacuum and one photon states, we recover the result that ${N\brace k}_1=\binom{N}{k}$.  In other words, if one can only pick $r_i\in \{0,1\}$ the problem of finding the number of ways to obtain $\ket k$ from \eq{phi} reduces to the amount of ways that one can choose $k$ 1's from $N$ boxes.  This result is then in agreement with the findings of \cite{ralph}.  Generally speaking, in order for the resulting fidelity to be high we require that the average photon number satisfy the constraint $\bar n/N \ll d$.  If this is true, then the success probability $P_{suc}$ will also be high, and the need to actually detect photons in the $t$-modes and condition on the empty outcome can be neglected.
\begin{figure}[htp]
\centering
\includegraphics[scale=0.8]{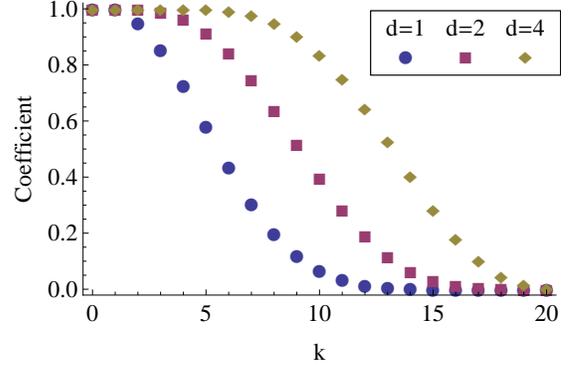}
\caption{Teleportation of Fock basis via N-splitters. To provide a fair comparison, $dN$ is fixed at a value of 20.  It is readily seen that the amplitudes associated with the teleported number states, given by \eq{trans}, remain higher for larger $d$.}
\label{fig:nkd}
\end{figure}
A simple way of observing how this procedure affects a state is to see how the coefficients in \eq{trans} scale as a function of the number of modes $N$ and the dimension of the maximally entangled states $d$.  In an ideal procedure we would have that for any $k$ the coefficient would be $1$, the actual results can be seen in \fig{nkd}.
\section{Entanglement Swapping}
We now turn to the specific case of teleporting one mode of an EPR state given by \cite{nielsen}
\begin{align}
\label{eq:epr}
\ket{\text{EPR}}&=\sqrt{1-\chi^2}\sum_{n=0}^\infty \chi^n\ket{n}_a\ket{n}_b,
\end{align}
where $\chi=\tanh(r)$, with $r\in [0,\infty)$ being the squeezing parameter.  The teleportation of one mode of an EPR pair models the teleportation of a Gaussian ensemble of different pure states, and thus this is an important case to study \cite{ralph}.  By applying equations \eq{tele} and \eq{psuc}, we see that an EPR state transforms as
\begin{align}
\ket{\text{EPR}_{tele}}&=\frac{\sqrt{1-\chi^2}}{\sqrt{P_{suc}}}\sum_{k=0}^{dN} \chi^n {N\brace k}_d\frac{k!}{N^k}\ket{k}_a\ket{k}_b\\
P_{suc}&=(1-\chi^2)\sum_{k=0}^{dN} \chi^{2k}{N\brace k}_d^2\frac{k!^2}{N^{2k}}\\
f&=\frac{1-\chi^2}{\sqrt{P_{suc}}}\sum_{k=0}^{dN}\chi^{2k}{N\brace k}_d\frac{k!}{N^k}
\end{align}
where $P_{suc}$ is the probability that no photons are detected in the $t$-modes, and $f$ is the fidelity between the input and output states.

\begin{figure}[htp]
\centering
\includegraphics[scale=0.65]{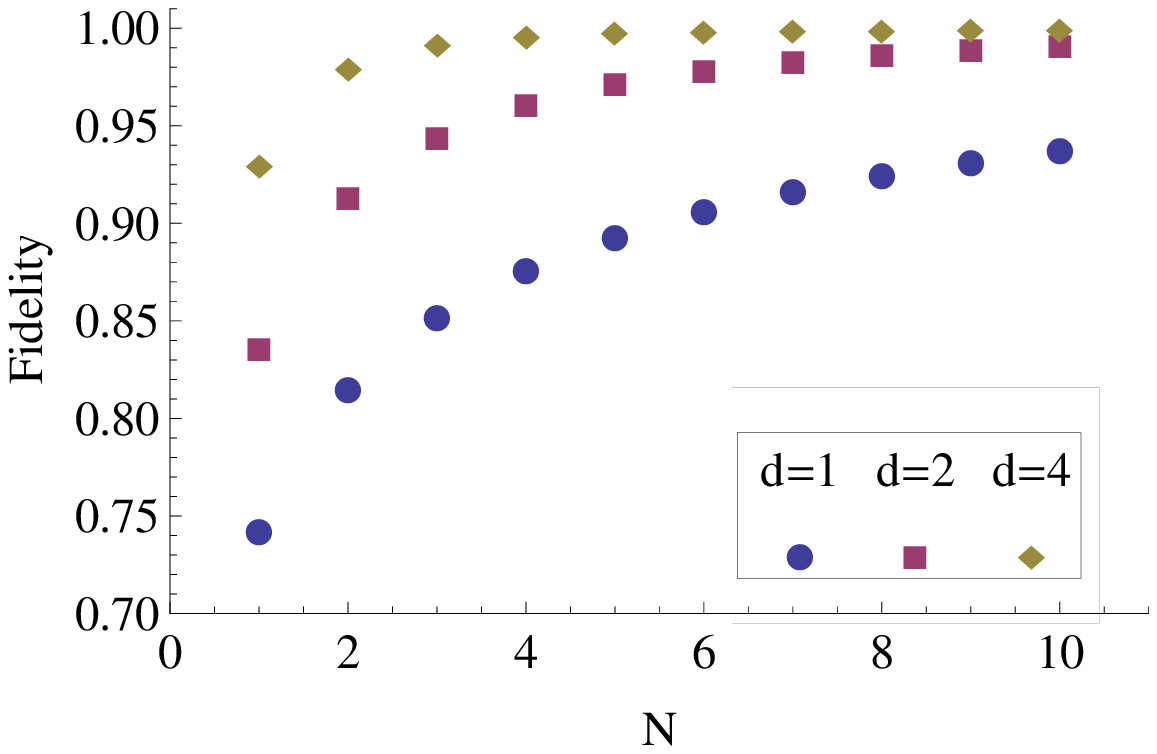}
\caption{Teleportation of one arm of an EPR pair with $V_s=10$.  Fidelity is plotted against the size of the $N$-splitter for a variety of different Hilbert space dimensions $d$ associated with each mode.}
\label{fig:fid}
\end{figure}
\begin{figure}[htp]
\centering
\includegraphics[scale=0.65]{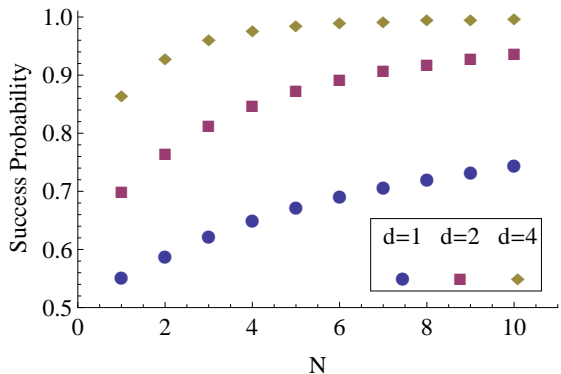}
\caption{Teleportation of one arm of an EPR pair with $V_s=10$.  Success probability is plotted against the size of the $N$-splitter for a variety of different Hilbert space dimensions $d$ associated with each mode.}
\label{fig:psuc}
\end{figure}
One measure of the amount of entanglement in an EPR pair is related to the degree of two-mode squeezing given by $V_s=(1+\chi)/(1-\chi)$, we can use this measure to observe how the fidelity and success probability vary as the number of entangled pairs is changed in \fig{fid} and \fig{psuc}.  By using larger Hilbert spaces, of dimension $d+1$, to teleport our CV state we are able to achieve comparable fidelities with fewer modes, as should be expected.  The usefulness of this result hinges on the ability to create maximally entangled states of a dimension greater than two, and while this problem has been studied \cite{max1,max2,max3} its practical implementation will likely require more experimental effort in this field.  The original result of Andersen and Ralph, the two-dimensional case, has the advantage of being practically feasible with today's technology since maximally entangled states in two-dimensions can be generated with a single photon source and an ideal 50/50 beamsplitter.  However, by moving to a higher dimension as proposed in \sii one can obtain a large increase in fidelity for modest values of $d=2,4$, even while using less modes $N'=d/N$ in the teleportation.

In general, it is found that when the input state has a low average energy this protocol is capable of outperforming the standard CV teleportation procedure.  Another interesting possibility is to use an alternative teleportation procedure for each mode after the first $N$-splitter.  For example, port-based teleportation \cite{asymp,port,recycle} has found application to programmable quantum processors, non-local quantum computation, and position-based cryptography \cite{crypt}.  As such, the ability to teleport a CV state without the need to apply a recovery operation directly on the system may prove to be a useful primitive in CV quantum computation.
\section{Noise and Loss}
An important consideration is that of the loss and imperfection inevitably associated with any physical realization of this protocol.  One cause for concern may arise from the fact that the quantum channel between Alice and Bob, to be used for the qudit teleportation, may not be maximally entangled.  If we model the deviation from being a pure maximally entangled state by the action of a depolarizing channel we obtain the state $\rho = p\ket\psi\bra\psi + (1-p)\hat{\mathcal I}/d^2$, where $\ket\psi$ is as given in \eq{psi}.  In this case it has been shown \cite{horodecki, tele2} that the channel fidelity is given by $f=p+(1-p)/d$.  This statement is equivalent to saying the fidelity is given by $f=(Fd+1)/(d+1)$, where $F$ is the maximal singlet fraction attainable through trace preserving local operations and classical communication.

One method to create such maximally entangled states involves distilling entanglement out of squeezed states \cite{max2}.  In general, the types of loss will depend on which scheme is utilized to generate a maximally entangled state.  In this procedure, a detailed discussion of the types of noise is given by Duan et al. \cite{duan}.  It is interesting to note that some forms of noise depend in one way or another on the dimension $d$ of the maximally entangled state involved, while others are independent of this factor.  The authors show that for the case of small preparation noise, the probability of distilling a maximally entangled state of dimension $d$ picks up an extra factor of $p_d\propto e^{(-\eta_A+\eta_B)\tau d}$ where the $\eta$'s are cavity loss factors and $\tau$ is a transmission time.  So preparation noise makes it less likely to obtain a maximally entangled state of higher dimension, but in the end one does obtain a pure state.  There are also other proposed methods of truncating CV states through teleportation \cite{trunc} that sidestep the requirement of starting with a maximally entangled state.

Another source of loss comes from imperfect detection as well as inaccurate photon counting.  Photon counting in a mode $\hat a$ by an imperfect detector can be modelled as a positive-operator-valued-measure (POVM) described by \cite{detection}

\begin{align}
\hat \Pi_N^{(a)}&=\sum_{n=0}^N\sum_{m=n}^\infty \frac{e^{-\nu} \nu^{N-n}}{(N-n)!}\eta^n(1-\eta)^{m-n}\binom{m}{n}\ket m\bra m,
\end{align}
where the elements sum to the identity.  The number of photons registered by the detector is given by $N$ whereas $n$ corresponds to the number of photons that actually enters the detector.  The detection efficiency is given by $\eta$ and the mean dark count rate is described by $\nu$.  Conventional avalanche photo-diodes can be described by a POVM $\{\hat \Pi_0^{(a)},\hat{\mathcal I}-\hat \Pi_0^{(a)}\}$ where the detector is only capable of distinguishing between no photons, and at least one photon.  Single photon resolving detectors are similarly described by a POVM with elements corresponding to $\{\hat \Pi_0^{(a)},\hat \Pi_1^{(a)},\hat{\mathcal I}-\hat \Pi_1^{(a)}-\hat \Pi_0^{(a)}\}$, and this generalizes in a straightforward manner to detectors capable of distinguishing up to an arbitrary number of photons.

Generally speaking the protocol will perform better when less terms are lost in the truncation, however this will typically come at a cost.  Either generation of maximally entangled states of higher dimension will succeed with lower probability or will result in a lower fidelity.
\section{Comparison to other schemes}
To consider the possible advantages and disadvantages of this new scheme, it will be useful to compare it with other possibilities of teleporting a CV quantum state.  The traditional approach to CV teleportation is to use a squeezed vacuum state of the form given by \eq{epr}.  However, the resulting fidelity generally scales exponentially with the amount of squeezing.  For example, the fidelity for teleportation of a coherent state is given by $F=1/(1+e^{-2r})$ where $r$ is the amount of squeezing.  In this case, the fidelity approaches unity only in the limit of an unphysical amount of squeezing, and thus this scheme does not scale well.

In an ideal case, both the scheme proposed by Andersen and Ralph \si as well as the scheme in this paper \sii will approach perfect fidelity with finite resources.  However, the possible advantages or pitfalls will lie in the physical implementations, and by taking account of imperfections.  Perhaps the most obvious source of loss is that of finite detector efficiency, so we will consider how the schemes perform while taking this into account.  Unfortunately, it becomes difficult to compare the two schemes on a level playing field since \si deals with qubits which can be implemented with linear optics and \sii requires nonlinearities.  Specifically, one must be able to implement the XOR gate \eq{xor}, and one must also be able to make measurements that distinguish between different photon numbers.  One possible implementation of the XOR gate is based on the Kerr effect \cite{gxor}, and phonton number resolving (PNR) detectors fulfil the second requirement.  However, the relatively low nonlinear susceptibility constants $\chi^{(3)}$ available in state-of-the-art materials leads to impractically long interaction times, and conventional qudit teleportation of this sort is not currently possible in the lab.

If we suppose that such a XOR gate could be implemented in an ideal fashion then we could compare \si and \sii by simply comparing the detection efficiencies.  In the case of \si the overall success rate would scale as $(1/2)^N$, where $N$ is the number of modes, if one were restricted to linear optics, since only two of the four Bell states can be distinguished.  To make a fair comparison to the qudit case, we assume that all of the generalized Bell states can be deterministically detected, in this case the detection efficiency of \si is given by $\eta^N\eta^N=\eta^{2N}$, where one factor of $\eta$ comes from the teleportation while the other comes from conditioning on no photons in the auxiliary modes.  Similarly, the detection efficiency of \sii would be given by $\xi_M^N\eta^N$, where $\xi_M$ is the effective detection efficiency of a PNR detector when distinguishing between vacuum through $M$ photons.

Unfortunately, for any of the above to be a meaningful comparison we would have to have values for $\eta$ and $\xi$, as well as being able to implement a XOR gate.  Fortunately, this is not the only proposed method for teleporting a system on a finite dimensional Hilbert space.  For example, we can consider the multimode interferometer approach found in \cite{trunc}. 
\begin{figure}[htp]
\centering
\includegraphics{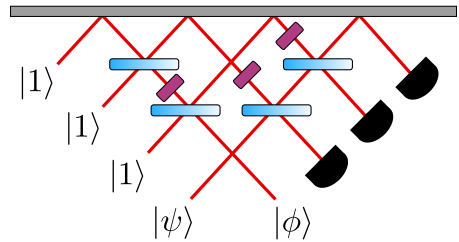}
\caption{Teleporting a quartit via multimode interferometer, image based off Figure 2 in Ref. \cite{trunc}.}
\label{fig:multimode}
\end{figure}
To take a specific case, consider teleporting a four dimensional qudit, in this case one can show that the probability of success can be made at least as high as $0.18$.  If one desires a fidelity of approximately $93\%$ using scheme \sii one could use either $N=11$ qubits or $N=3$ quartits, as can be seen in \fig{fid}.  The success probability using the qubits scales as $(1/2)^{11}\eta^{11}$ whereas it scales as $(0.18)^3(\xi_1^3)^3$ for the quartits; the factor of $0.18$ coming from the success probability of the interferometer post-selection.  One can distinguish between zero, one, and more than one photons using a single photon resolving detector, such as a visual light photon counter (VLPC) \cite{vlpc}, and this type of detector is sufficient to teleport a quartit state using the interferometer approach.  It is also possible to use this approach with a conventional photodetector, however one suffers a loss of fidelity.  An important remark is that for a reasonable range of values for $\eta$ and $\xi_1$ the proposed scheme \sii offers an advantage over the two-level scheme \si.  A general advantage of the new work detailed in scheme \sii is that one does not need to work with as many modes to achieve a larger, yet there are still unique difficulties in extending the dimension to include higher photon numbers.

\begin{figure}[Ht!]
\centering
\includegraphics[scale=0.85]{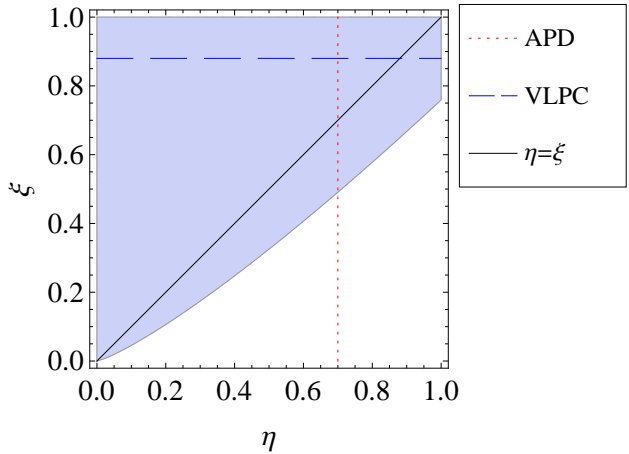}
\caption{Subset of parameter space where scheme \sii outperforms scheme \si in the specific case of teleporting quartits with a multimode interferometer approach.}
\label{fig:regionplot}
\end{figure}
Another potential method to teleport each mode after the $N$-splitter is to use a continuous variable EPR state to teleport a discrete variable state using gain tuning \cite{gaintune}.  This approach has the advantage of being deterministic, so although the fidelity will be degraded due to finite squeezing, the efficiency of the protocol may still be higher than the interferometer approach discussed above.  This approach has already been used to demonstrate teleportation of a dual-rail qubit with fidelity on the order of $\sim 0.8$ \cite{gain2}.
\section{Summary}
To conclude, we have demonstrated an extension to the CV teleportation scheme proposed by Andersen and Ralph \cite{ralph}.  This protocol splits an input CV state equally among $N$ modes, and then teleports each mode using a finite dimensional Hilbert space with $\text{dim}(\mathcal H)=d$.  The reduction of CV states to a $d$-dimensional space results in a truncation of the Fock basis, and corresponds to a loss in fidelity.  However, by choosing an appropriate number of modes $N$ and dimension $d$, one is able to achieve arbitrary fidelity with high probability.  The ability to create and distribute maximally entangled qudits of dimension $d$ is a technique required in order for this protocol to be practically feasible.  An interesting possibility opened up by this approach is to use generalized teleportation protocols for each mode after the input is split.  One is then able to harness the results for systems of qudits and attempt to apply them to the field of continuous variable quantum information.
\appendix*
\section{}
\label{sec:app}
To obtain the coefficient of a number state $\ket k$ in the expansion of the state
\begin{align}
\label{eq:a1}
\ket \phi &=\mathcal{N}\left(\hat{\mathcal I} + \frac{\alpha}{N}\hat a_f^\dagger +\ldots+ \frac{\alpha^d}{d!N^d}(\hat a_f^\dagger)^d\right)^{N}\ket{0}
\end{align}
consider all of the ways one can get a term $\hat a_f^\dagger$ to the power $k$, for a fixed but arbitrary $k$.  In expanding equation \eq{a1}, we take one term from each of the $N$ summations that are being multiplied.  For example, one term in the expansion corresponds to the set of values $r_1,r_2,\ldots,r_N$ where $r_i\in\{0,1,\ldots,d\}$ for all $i$.
\begin{align}
\prod_i^N \frac{\alpha^{r_i}}{r_i!N^{r_i}}(\hat a_f^\dagger)^{r_i}\ket 0
\end{align}
To find the coefficient $c_k$ of the number state $\ket k$, where $\ket \phi=\sum_{k=0}^M c_k \ket k$, we consider the sum of all such terms with the constraint that $\sum_i r_i=k$.  Recalling that $\ket k = (\hat a^\dagger)^k/\sqrt{k!}$ we have that
\begin{align}
c_k&=\sum_{\substack{\{r_1,r_2,\ldots,r_N\} \\ r_1+r_2+\ldots+r_n=k \\ r_i \leq d~ \forall i}} \prod_{r_j} \frac{1}{r_j!}\left(\frac{\alpha}{N}\right)^k \sqrt{k!},
\end{align}
which can be calculated numerically.

\end{document}